\documentclass[12pt]{article}
\newcommand\be{\begin{equation}}
\newcommand\ee{\end{equation}}
\newcommand\ba{\begin{eqnarray}}
\newcommand\ea{\end{eqnarray}}
\newcommand\eq{\begin{equation}}           
\newcommand\en{\end{equation}}

\input epsf
\usepackage{amssymb}
   \usepackage{epsfig,cite}
\include{epsf}

\usepackage{subfigure}

\begin{document}
\title{
{\hfill  \small     MCTP-09-50   \\~\\~\\}
Enhanced Tau Lepton Signatures at LHC \\
in Constrained Supersymmetric Seesaw
\author{Kenji Kadota$^1$ and Jing Shao$^{1,2}$  \\
{\em \small 
$^1$ Michigan Center for Theoretical Physics, University of Michigan, Ann Arbor, MI 48109} \\
 {\em \small
$^2$ Department of Physics, Syracuse University, Syracuse, NY 13244}
}}
\maketitle   
\begin{abstract}
We discuss the possible enhancement of the tau lepton events, 
with the emphasis on the signals with three or more hadronic taus, at LHC when the left-handed stau doublet becomes light or even lighter than the right-handed one. This is the natural outcome of the supersymmetric seesaw models where the slepton doublet mass is suppressed by the effects of a large neutrino Yukawa coupling. We study a few representative parameter sets in the sneutrino coannihilation regions where the tau sneutrino is next-to-lightest supersymmetric particle (NLSP) and the stau coannihilation regions where the stau is NLSP both of which yield the thermal neutralino LSP (lightest supersymmetric particle) abundance determined by WMAP.

\end{abstract}

\setcounter{footnote}{0} 
\setcounter{page}{1}
\setcounter{section}{0} \setcounter{subsection}{0}
\setcounter{subsubsection}{0}

\section{Introduction}
Supersymmetry (SUSY) is one of the most promising candidates beyond the Standard Model (SM).  
Implementing the seesaw mechanism \cite{seesaw} in SUSY is also well motivated from the finite neutrino mass and the implications of the supersymmetric seesaw such as the lepton flavor violations and the dark matter constraints have been studied \cite{borzu,hisa,hisa2,casa3,casa,how4,baer2,Lav,el,bla,fre,hitoshi,gordy3,hints,de,dep,iba,barg1,hir,nucmssm,est,snu3,tao,barg2}. We in this paper pay particular attention to the suppression of the mass of left-handed stau doublet $(\tilde{\nu}_{\tau},\tilde{\tau}_L)$ due to the RGE (renormalization group equation) effects of an order unity neutrino Yukawa coupling in the type-I seesaw scheme and discuss its consequences on the inclusive tau signatures at LHC. The main result is a large number of 3 or more hadronic $\tau$ events. 

Even though the tau identification and reconstruction are harder than 
those of electron/muon, $\tau$ events can play an important role in probing the underlying theory beyond the SM. 
Taus decay mostly hadronically ($\sim 65$ \%) in which about $50$\% result in one charged pion (single-prong decays) and the rest in three charged pions (three-prong decays) (plus neutral particles $\nu_{\tau}$ and 
$\pi^0$'s), and their hadronic jets are characteristic for their high collimation and low multiplicity.
This is in contrast to the leptonic tau decays which involve two neutrinos and tend to give rise to relatively soft leptons ($e$, $\mu$), and it is hard to distinguish them from the primary leptons.
We in this paper consider only the dominant hadronic $\tau$ decays and taus hereafter refer to the hadronically decaying taus while leptons refer to $l=e,\mu$ distinguished from $\tau$.

The collider studies for the tau events in a light stau scenario have been discussed for the constrained minimal supersymmetric standard model (CMSSM) in the context of 
a large $\tan \beta$ or a non-vanishing $A_0$ \cite{nojidrees,bae2,drees,kao2,joe,james,matchepie,mat1,hinch2,mre,hinch,dark1,nonzeroA,nonzeroA2,bha,shafi,ara}. In CMSSM, the right-handed stau is typically lighter than the left-handed stau doublet and a large left-right mixing due to 
the large $\tan \beta$ or $A_0$ helps the lighter stau mass eigenstate $\tilde{\tau}_1$ to have a smaller mass and more $SU(2)_L$ component.
In the seesaw scheme, however, a neutrino Yukawa coupling can suppress the mass of stau doublet (which can be even below the right-handed stau mass) 
so that a light $\tilde{\tau}_1$ with a significant left-chiral component even without any left-right mixings can be easily achieved. This is essentially the reason for the enhancement of the tau signatures at LHC.

\S \ref{twomod} qualitatively outlines the difference between a conventional light 
right-handed stau scenario in CMSSM and a light left-handed stau scenario in the constrained supersymmetric seesaw. We quantitatively discuss, in \S \ref{eg}, the effects of a large neutrino Yukawa coupling on the (3 and 4) tau signals at LHC for a few benchmark points in the sneutrino coannihilation \cite{nucmssm,snu3} and stau coannihilation regions  \cite{stau}. Fig.~\ref{brfigure} shows the enhancements (suppressions) of the branching ratios of the light chargino decay to $\tau$ ($e/\mu$) and 
Fig.~\ref{number} shows the 3 tau event enhancements observable at LHC. The brief discussions for other channels ($2\tau+1l$,$1\tau+2l$ and $3l$ signals) in the seesaw scenario are given in \S\ref{dis} followed by the conclusion.

\section{Model}
\label{twomod}
\subsection{CMSSM}
\label{mod}
Before presenting the constrained supersymmetric seesaw scenarios which can naturally realize the light left-handed stau doublet, let us briefly review the CMSSM scenarios (often using a large $\tan \beta$) where the light stau is right-handed dominated for the comparison between these two scenarios.

In CMSSM or mSUGRA (minimal supergravity), the right-handed sleptons are typically lighter than the corresponding left-handed sleptons because the left-handed ones are pushed up from $SU(2)_L$ weak couplings in addition to U(1) hypercharge couplings in their renormalization group (RG) evolutions. Among the right-handed sleptons, $\tilde{\tau}_{R}$ is typically lighter than 
$\tilde{e}_R/\tilde{\mu}_R$ because of the larger Yukawa coupling driving it down in the RG evolutions. Note the second lightest neutralino and the lightest chargino $\chi^0_2, \chi^{\pm}_1$ (which many SUSY cascade decays go through before reaching the lightest neutralino LSP) are typically the superaprtner of $SU(2)_L$ gauge boson $W^3,W^{\pm}$ in the CMSSM (as well as in 
the seesaw model to be discussed). Hence the decays of $\chi^{\pm}_1,\chi^0_2$ into $\tilde{\tau}_1$ are expected to be suppressed when 
$\tilde{\tau}_1$ is dominantly right-handed. For these reasons, $\tau$ signatures from the light stau have been often emphasized in the context of a large $\tan \beta$ which can increase the stau mass off-diagonal term for the large left-right mixings \cite{bae2,drees,kao2,joe,james,matchepie,mat1,hinch2,mre,hinch}. It also enhances the tau Yukawa coupling ($\propto \tan \beta$) to drive the stau mass lighter. A large $\tan \beta$ however affects other physical observables such as $Br(b \rightarrow s \gamma)$ and $Br(B_s\rightarrow \mu^+ \mu^-)$ which tightly constrain the kinematically preferable small $m_0,m_{1/2}$ parameter regions. 

These conventional pictures for the light stau totally change in the seesaw scenario and the small $m_0,m_{1/2}$ regions now open up which can lead to the enhanced tau events at a collider and also to the desired dark matter relic abundance.

\subsection{Constrained Supersymmetric Seesaw}
\label{model}
We here briefly outline the constrained supersymmetric seesaw model which can realize the light left-handed slepton doublet.
Seesaw mechanism is well motivated to explain a tiny neutrino mass, and we make a minimal extension of CMSSM by introducing a heavy right-handed neutrino whose superpotential reads
\ba
W=W_{CMSSM}+y_N N^c L H_u+\frac{1}{2}M_N N^c N^c
\ea
where $N$ is a right-handed neutrino with the mass $M_N$ and $y_N$ is a neutrino Yukawa coupling. 
For simplicity, we consider one heavy right-handed neutrino in the third generation ignoring the flavor mixings and CP phases in the neutrino sector. Our model is `constrained' because we still keep the universal boundary conditions for the soft SUSY breaking terms such as the 
scalar mass $m_N$ and the trilinear coupling $A_N$. The constrained seesaw model therefore has, in addition to the conventional CMSSM parameters
\ba
m_0,M_{1/2},A_0,\tan \beta,sign(\mu)
\ea
the neutrino mass parameters
\ba
M_N(Q_{GUT}),m_{\nu}(Q_{m_Z})
\ea
The heavy right-handed neutrino mass $M_N$ and the light neutrino mass $m_{\nu}$ 
are specified respectively at the GUT scale and at the Z boson mass scale $m_Z$ (we are interested in $M_N$ less than the grand unification (GUT) scale).

A right-handed neutrino decouples at $M_N$ which itself 
runs from the GUT scale down to $M_{N}$. Below the energy scale $M_{N}$, 
$N$ is integrated out and the effective Lagrangian includes the residual 
higher order operators suppressed by $M_N$. Because we are interested in
 the value of $M_{N}$ relatively close to the GUT scale for an order unity $y_N$, those non-renormalizable operators 
do not affect the mass spectrum at the low energy scale other than a left-handed neutrino mass which receives a dominant contribution from the dimension five operator
\ba
L_5  \ni -\kappa (L H_u)(L H_u)
\ea 
We hence keep the RGE of $\kappa$ in our 
full two-loop RGEs and $\kappa \langle H_u \rangle^2 $ 
is matched to $m_{\nu}(Q_{m_Z})$ \cite{ant,ant2}. Even though we numerically solve the full two-loop RGEs including the neutrino parameters, it would be instructive to list here the analytical form of the one-loop RGE for the tau slepton doublet
\ba
\frac{d}{dt}m_{L_3}^2&=&\frac{1}{16 \pi^2}\left(2 y_{\tau}^2X_{\tau}+2 y_N^2X_N-\frac65g_1^2 M_1^2-6g_2^2 M_2^2-\frac{3}{5}g_1^2 S\right)+... \label{rge1}
\ea
where the $y$'s are Yukawa couplings, $g$'s are gauge couplings,  $t=\log Q$ and 
\ba
X_\tau&=&m_{L_3}^2 + m_{\tau_R}^2 + m_{H_d}^2 + A_\tau^2\\
X_N&=&m_{L_3}^2+m_{{N}}^2+m_{H_u}^2+A_N^2 \\
S&=&Tr(m_{ Q}^2 + m_{ D}^2 -2 m^2_{ U}
-m_{ L}^2 +m_{E}^2)
+ {m}^2_{H_u}-{m}^2_{H_d}
\ea
In the constrained seesaw as well as in CMSSM, $S=0$ at the GUT scale due to the universality of the soft scalar masses. Deviations from $S=0$ are due to the RGE evolutions at the two-loop level and hence $S$ does not play a significant role for our study.
We here can explicitly see that the neutrino Yukawa coupling can drive down the tau slepton doublet mass for a large $y_N^2 X_N$. A small $m_0$ would be preferable for the energy reach at LHC, and we choose in turn a large $A_0$ to obtain a large $X_N$. A byproduct of our preferring a large (negative) $A_0$ for a bigger effect of the neutrino Yukawa coupling is that a non-vanishing trilinear scalar parameter can also increase the Higgs mass due to the loop contributions of the lighter stop because of the left-right mixing in the
stop mass matrix. Choosing $A_0$ excessively large, however, causes the Higgs mass to decrease and also leads to tachyonic squark/slepton masses. In fact, the vanishing $A_0$ is often used in the study of CMSSM and the small $m_0,m_{1/2}$ regions are disfavored by the Higgs mass bound $m_h > 114~$GeV \cite{lepa}. Our choice of a relatively large $A_0$ alleviates
this problem and the energetically accessible $m_0,m_{1/2}$ parameter space now opens up.

We point out the left-handed stau doublet can even become lighter than the right-handed stau, so that the lighter stau mass eigenstate $\tilde{\tau}_1$, in contrast to CMSSM, can become left-handed dominated as shown in a concrete example in the next section (for instance in Fig.~\ref{rgefig}). In such a case, the interactions of $\tilde{\tau}_1$ can be increased by the 
$SU(2)_L$ gauge interactions and it is natural to 
expect the enhanced tau events via the light $\tilde{\nu}_{\tau}$ and $\tilde{\tau}_1$ coming from the decays of $\chi^0_2$ and $\chi^{\pm}_1$ which are typically wino-like. In contrast to CMSSM, a small $\tan \beta$ would be preferable to obtain the 
light left-handed stau doublet because $y_N\propto \cot \beta$, which 
also helps in keeping the $SU(2)_L$ nature of $\tilde{\tau}_1$ by suppressing the left-right mixings when the left-handed stau doublet becomes lighter than the right-handed stau. A small $\tan \beta$ is also advantageous to relax the other physical constrains such as $Br(B_s\rightarrow \mu^+ \mu^-)$.

\section{Examples}
\label{eg}

\subsection{Model Points}

We here illustrate our discussions in the last sections more quantitatively for a few benchmark points. We, among the possible vast parameter sets, consider the parameter regions which can give 
the thermal relic abundance of the neutralino LSP in the desired WMAP (95\% CL) range, $0.0975 <\Omega_{DM} h^2<0.1223$ \cite{wmap}. Among such cosmologically motivated parameter regions which now give the significantly reduced parameter space, the small $m_0,m_{1/2}$  
would be preferred for their kinematical reach at LHC and we choose the
representative sets of parameters with the small $m_0,m_{1/2}$ for the sneutrino coannihilation regions \cite{nucmssm,snu3} where the tau sneutrino $\tilde{\nu}_{\tau}$ is NLSP and the stau coannihilation regions where the stau $\tilde{\tau}_1$ is NLSP \cite{stau}. Both sneutrino and stau coannihilation regions show up for the small $m_0,m_{1/2}$ with 
the sneutrino coannihilation regions corresponding to the smaller $m_{1/2}$ and bigger $m_0$ than 
those in the stau coannihilation regions. Note, if one uses a large $\tan \beta$ as commonly discussed in the previous literature on the light stau scenarios in CMSSM, 
the small $m_0,m_{1/2}$ giving the desired thermal relic abundance are disfavored by, for instance, $Br(b \rightarrow s \gamma)$.

Table \ref{bench1} lists our choices of the parameters for which the collider events were generated by PYTHIA \cite{code2}.
Point A represents the sneutrino coannihilation regions where $y_N(Q_{GUT})\sim 1.8,y_N(M_N)\sim 1.5,M_N(Q_{GUT})=8 \times 10^{14},M_N(M_N)\sim 6\times 10^{14}~$GeV, and Point B is in the stau coannihilation regions where $y_N(Q_{GUT})\sim 1.5,y_N(M_N)\sim 1.3,M_N(Q_{GUT})=6 \times 10^{14},M_N(M_N)\sim 5 \times 10^{14}~$GeV. The corresponding mass spectra and the physical observables are given in Table \ref{massspec}.  The seesaw and CMSSM at each parameter set share the common conventional universal values $m_0,m_{1/2},A_0,\tan \beta, sign (\mu)$. The mass spectra were calculated by using the modified SUSPECT incorporating a heavy right-handed neutrino at the two loop level based on the descriptions in \S \ref{model} and the physical observables (the dark matter abundance $\Omega_{DM}h^2$, the branching ratios of $b \rightarrow s \gamma,B_s \rightarrow \mu^+ \mu^-$ and the deviation of the anomalous 
magnetic moment of muon from SM prediction $a_{\mu}=(g-2)_{\mu}/2$) were calculated by MicrOMEGAs and the total 
inclusive cross section $\sigma_{tot}$ from PYTHIA \cite{codes}. The total inclusive cross section is larger for Point A because of a small $m_{1/2}$ which reduces the gaugino masses (in particular the gluino mass) and also suppresses the RG evolutions of the sparticles by the gauge interactions.
For the 
seesaw scenario which gives the desirable thermal relic abundance for these two benchmark points, we checked our choices of the parameters are consistent at the 3$\sigma$ level with the constraints coming from the $BR(b \to s \gamma)=(3.55\pm 0.24^{+0.12}_{-0.13})\times 10^{-4}$ \cite{bsg1,bmm}, the experimental upper bound $BR(B_s \rightarrow \mu^+ \mu^-)<4.7 \times 10^{-8}$ \cite{cdf,bmm} and $a_{\mu}=(g-2)_{\mu}/2=(30.2\pm 8.8)\times 
10^{-10}$ \cite{g2}. The CMSSM cases at these benchmark points are studied for the purpose of the comparison with the seesaw scenario. We shall show the big enhancement in the tau signal events in the seesaw scenario compared with CMSSM.

\begin{figure}
             \centering
                 \includegraphics[width=0.47\textwidth]{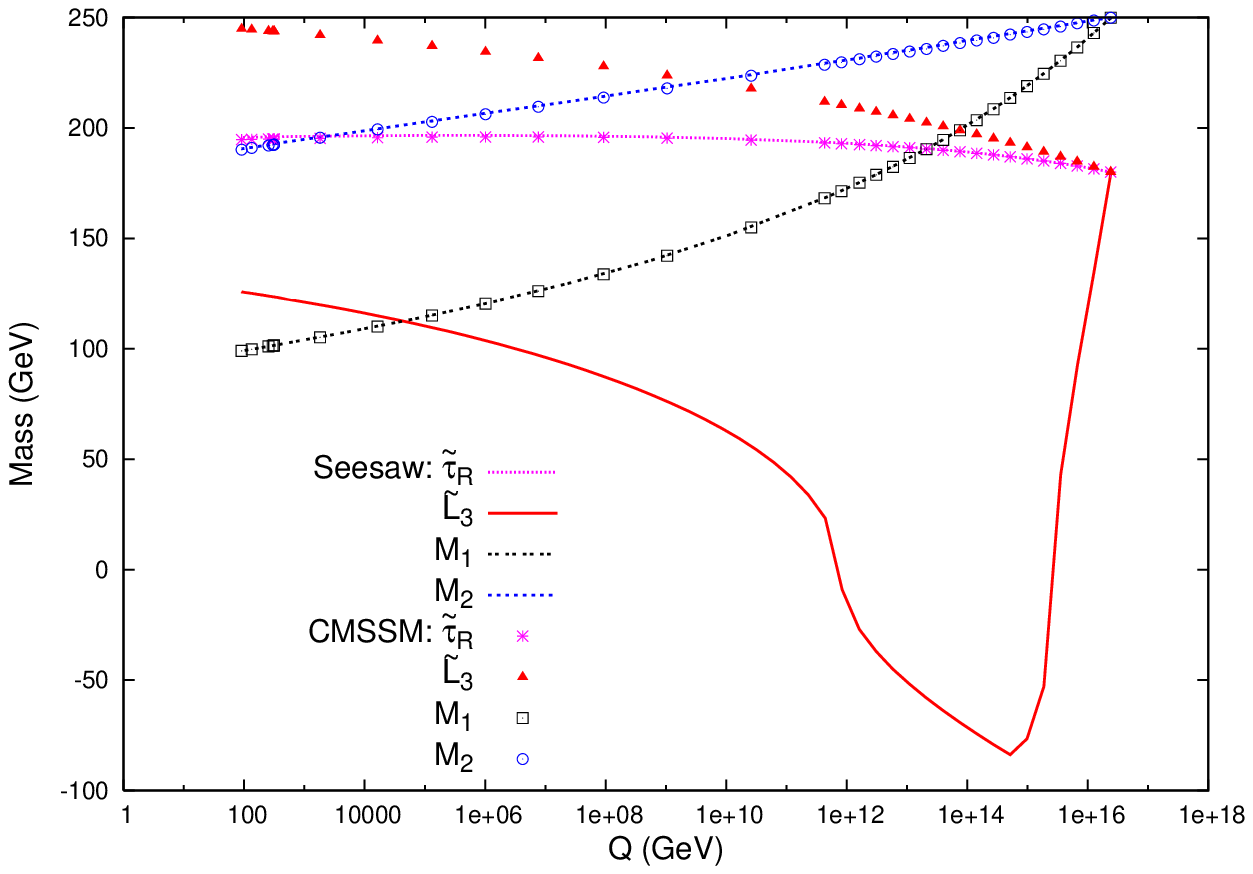}
                 \includegraphics[width=0.47\textwidth]{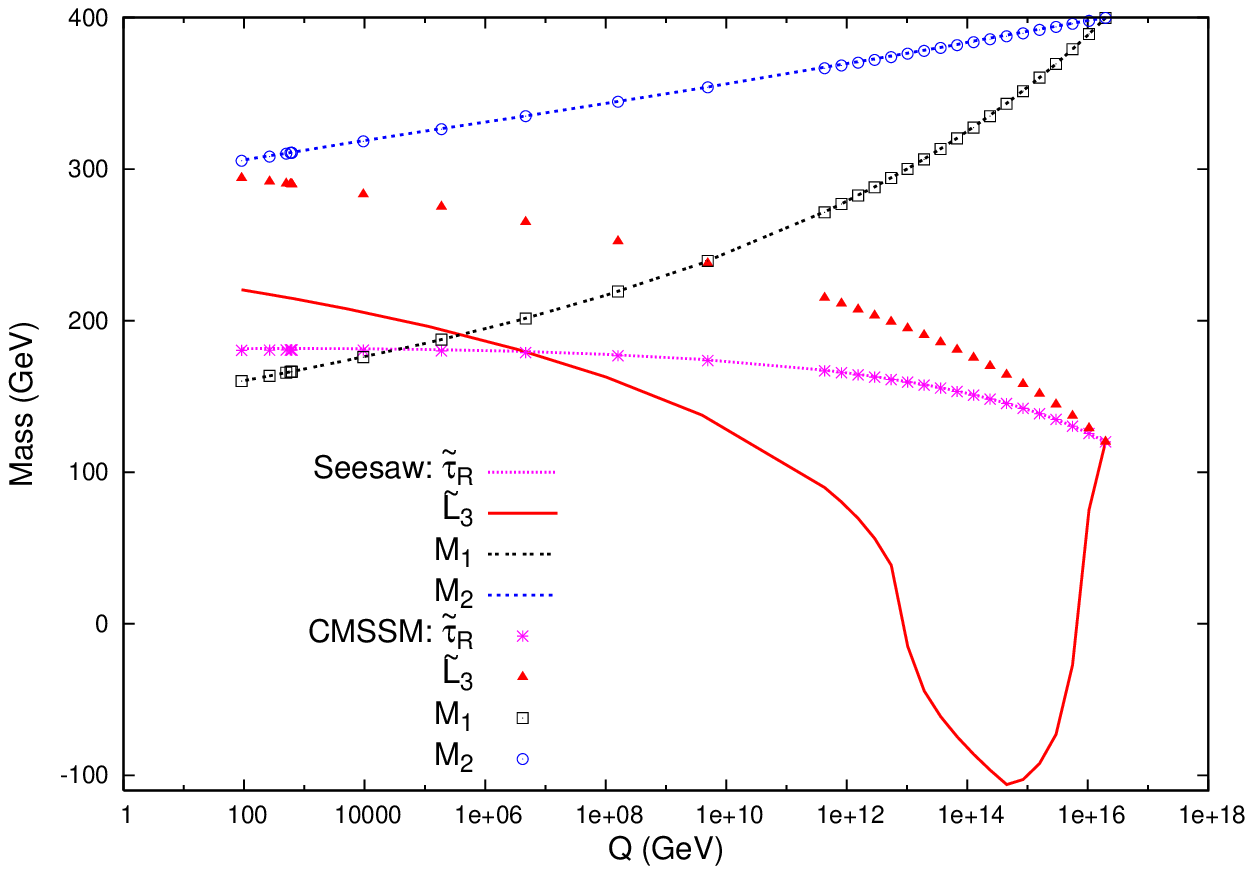}
 \caption{RG evolutions of the right-handed stau, stau doublet and gaugino masses for our benchmark points A (left) and B (right). The mass differences between the seesaw and CMSSM for the right-handed stau and the gauginos are at the two-loop level and hence not visible in this figure.}
     \label{rgefig} 
     \end{figure}

Fig.~\ref{rgefig} shows the RG evolutions of the masses for these two benchmark mark points which illustrate our discussions in \S \ref{model} ($M_1,M_2,L_3$ in the figure represent respectively the first and second generation gauginos and stau doublet). Note the slepton doublet is affected by a neutrino Yukawa coupling at the one loop level while the other masses shown in this figure are affected at the two loop level (hence those mass differences between the seesaw and CMSSM are not visible in the figure). We can see the slepton doublet mass is driven down by 
the neutrino Yukawa coupling $y_N$ from the GUT scale down to the right-handed neutrino mass scale $M_N$ at which the right-handed neutrino decouples.

\begin{table}
\centering
\begin{tabular}{|c|c|c|}
 \hline  
  & Point A  & Point B \\ 
   & (sneutrino coannihilation region) & (stau coannihilation region) \\ \hline
 $m_0$  & 180 GeV &  120 GeV \\ \hline
$m_{1/2}$ &  250 GeV  &  400 GeV \\ \hline
 $A_0$  & -900 GeV &  -900 GeV \\ \hline
$\tan \beta$ &  6.5  &  7 \\ \hline
 $sign(\mu)$  & 1 &  1 \\ \hline
$M_N(Q_{GUT})$ &  $8 \times 10^{14}~$GeV &  $6 \times 10^{14}~$GeV \\ \hline
 $m_{\nu}(Q_{m_Z})$  & 0.05 eV & 0.05 eV\\ \hline
\end{tabular}
\caption{The benchmark points for the sneutrino (Point A) and stau coannihilation (Point B) regions. $M_N$ and $m_{\nu}$ are the values specified at the GUT scale and at $m_Z$ respectively.}
\label{bench1}
\end{table}

\begin{table}
\begin{minipage}
{0.5\linewidth}\centering
\begin{tabular}{|c|c|c|}
 \hline  
 Point A & Seesaw & CMSSM \\ \hline 
$\tilde{\tau}_1$ &  121 &  193 \\ \hline
 $\tilde{\tau}_2$  & 207 &  253 \\ \hline
$\tilde{\nu}_{\tau}$ &  106  &  236 \\ \hline
$\tilde{l}_l$ &  251  &  251 \\ \hline
 $\tilde{l}_R$  & 207 &  207 \\ \hline
$\tilde{\nu}_{l}$ &  239  &  239 \\ \hline
 ${\chi}^0_1$  & 99 &  99 \\ \hline
${\chi}^0_2$ &  191  &  190 \\ \hline
${\chi}^{\pm}_1$ & 191  &  190 \\ \hline
$h$ &  114  &  114 \\ \hline
$\tilde{t}_1$ &  259  &  175 \\ \hline
 $\tilde{b}_1$  & 498 &  480 \\ \hline
 $\tilde{g}$  & 611 &  611 \\ \hline
$\Omega_{DM}h^2$ &  0.099 &  0.40 \\ \hline
$Br(b \rightarrow s \gamma)$ & $2.9 \times 10^{-4}$ & $2.4 \times 10^{-4}$ \\ \hline
$Br(B_s \rightarrow \mu^+ \mu^-)$ & $3.1 \times 10^{-9}$ & $3.1 \times 10^{-9}$ \\ \hline
$ a_{\mu}$ &  $14 \times 10^{-10}$  & $14 \times 10^{-10}$  \\ \hline
 $\sigma_{tot}(pb)$  & 49 &  119 \\ \hline
\end{tabular}
\end{minipage}
\hspace{0.5cm}
\begin{minipage}
{0.5\linewidth}
\centering
\begin{tabular}{|c|c|c|}
 \hline  
 Point B & Seesaw & CMSSM \\ \hline 
$\tilde{\tau}_1$ &  171  &  180 \\ \hline
 $\tilde{\tau}_2$  & 232 &  297 \\ \hline
$\tilde{\nu}_{\tau}$ &  205  &  283 \\ \hline
$\tilde{l}_l$ &  297  &  296 \\ \hline
 $\tilde{l}_R$  & 195 &  195 \\ \hline
$\tilde{\nu}_{l}$ &  286  &  286 \\ \hline
 ${\chi}^0_1$  & 163 &  163 \\ \hline
${\chi}^0_2$ &  312  &  312 \\ \hline
${\chi}^{\pm}_1$ &  312  &  312 \\ \hline
$h$ &  116  &  116 \\ \hline
$\tilde{t}_1$ &  531  &  499 \\ \hline
 $\tilde{b}_1$  & 761 & 751 \\ \hline
 $\tilde{g}$  & 934 &  934 \\ \hline
$\Omega_{DM}h^2$ &  0.11  &  0.25 \\ \hline
$Br(b \rightarrow s \gamma)$ &$3.5 \times 10^{-4}$&$3.4\times 10^{-4}$ \\ \hline
$Br(B_s \rightarrow \mu^+ \mu^-)$ &$3.1 \times 10^{-9}$&$3.1 \times 10^{-9}$\\ \hline
$ a_{\mu}$ &  $11 \times 10^{-10}$&$ 11 \times 10^{-10}$ \\ \hline
 $\sigma_{tot}(pb)$  & 4.3  &  4.3 \\ \hline
\end{tabular}
\end{minipage}
\caption{The mass spectra (in GeV) and the physical observables at Points A (left) and B (right) corresponding to Table \ref{bench1}.
The seesaw and CMSSM at each point share the common conventional universal values $m_0,m_{1/2},A_0,\tan \beta, sign (\mu)$.}
\label{massspec}
\end{table}
            \begin{figure}
             \centering
                 \includegraphics[width=0.47\textwidth]{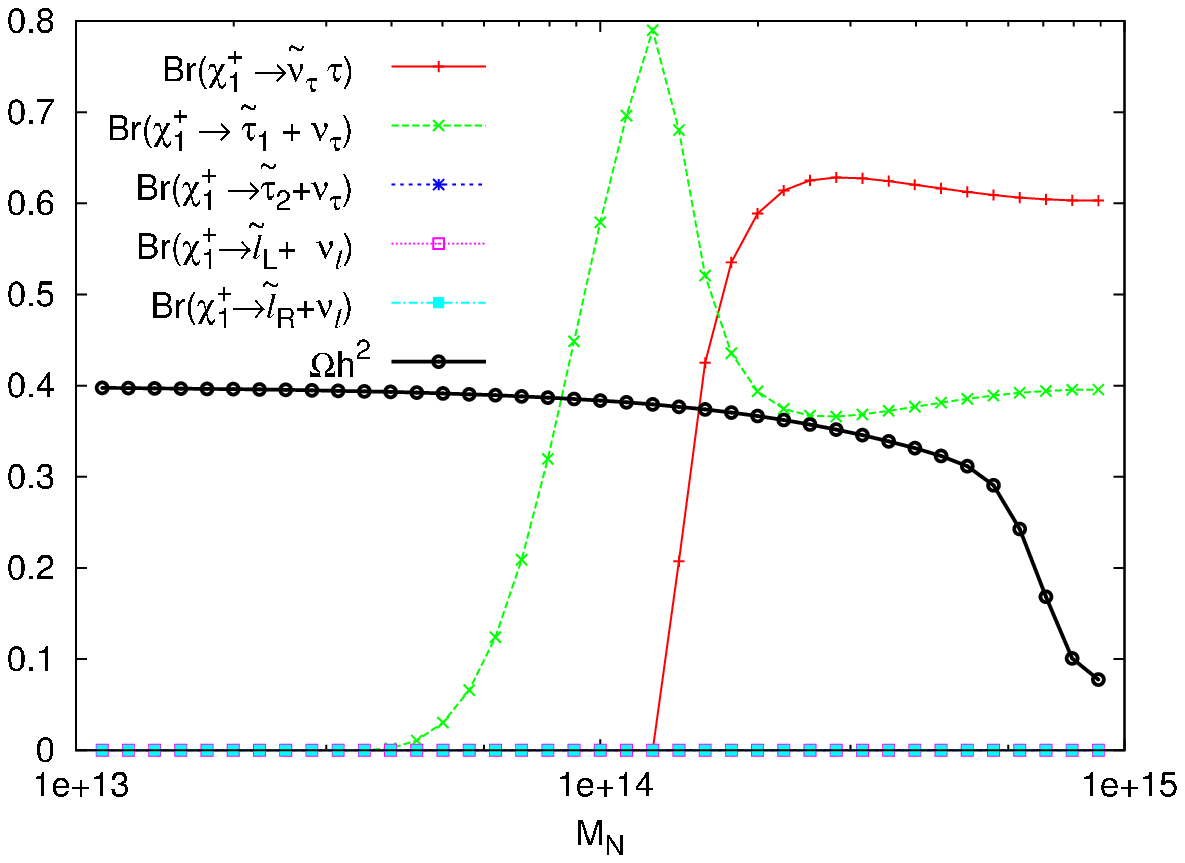}
                 \includegraphics[width=0.47\textwidth]{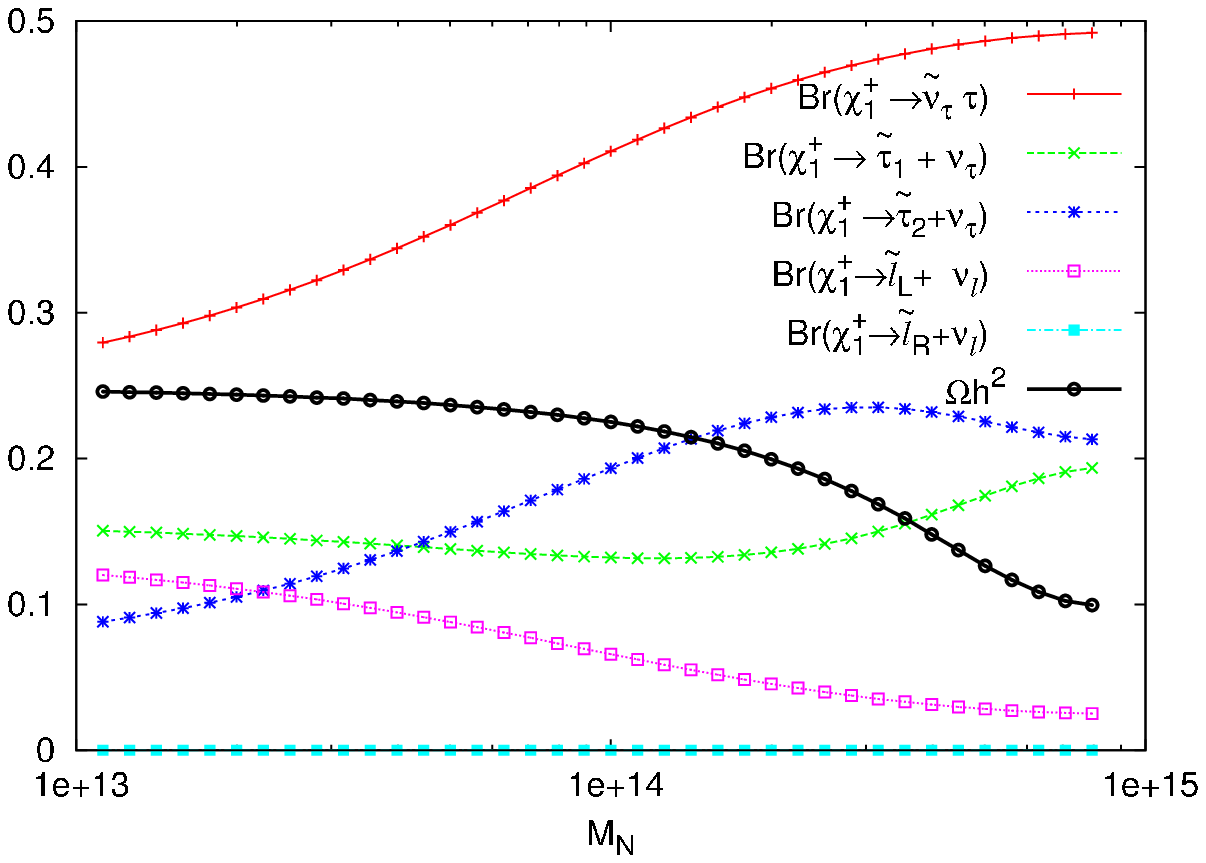}
 \caption{The branching ratios and the dark matter (neutralino LSP) abundance as a function of the right-handed neutrino mass $M_N~$(GeV) for our benchmark points A (left) and B (right).}
     \label{brfigure} 
     \end{figure}

Because many superparticle cascade decays go through the lightest chargino and the second lightest neutralino, the branching fractions of the $\chi^{\pm}_1,\chi^2_0$ play a significant role in determining the signals at LHC. Fig.~\ref{brfigure} shows the branching ratios of the lightest chargino into stau and slepton along with the neutralino dark matter abundance ${\Omega}_{DM}h^2$ as a function of the right-handed neutrino mass $M_N$ \footnote{A small $M_N$ limit essentially reproduces the CMSSM results because the neutrino Yukawa coupling becomes too small to affect the RG evolutions. Too large a $M_N$ beyond the range in this figure makes the neutrino Yukawa coupling too large and the 
perturbative calculations could become unreliable (for instance the two loop part of the beta function of $y_N$ becomes roughly a quarter of their one loop contributions for $y_N /16 \pi^2\sim  1/10$) \cite{casa3,bm,nume}.}. For Point A, the decays of $\chi^0_2$ and $\chi^{\pm}_1$ into stau and selectron/smuon are kinematically prohibited in the small $M_N$ region. When the right-handed neutrino becomes heavy enough, however, a large neutrino Yukawa coupling can drive the stau doublet mass down enabling $\chi^0_2,\chi^{\pm}_1$ to decay into the stau and tau sneutrino. The effects of a neutrino Yukawa coupling is evident for Point A in Fig.~\ref{brfigure} to result in the essentially 100\% branching ratio of $\chi^+_1$ into $\tilde{\tau}_1$ and $\tilde{\nu}_{\tau}$.  
For Point B, the decays of  $\chi_1^{\pm}$ and $\chi_2^0$ are slightly different, where their decay channels into all the stau/slepton species are kinematically allowed for both small and large $M_N$. In the right-panel of Fig.~\ref{brfigure}, for the sufficiently small $M_N$ (essentially CMSSM), the chargino decays into $\tilde{\tau}_1$ rather than into $\tilde{\tau}_2$ are preferred due to the available phase space while the chargino decays more preferentially into $\tilde{\nu}_{\tau}$ than into $\tilde{\tau}_1$ despite $m_{\tilde{\nu}_{\tau}}>m_{\tilde{\tau}_1}$. The latter occurs because the tau sneutrino is lighter than the left-handed stau due to the D-term contributions and the effects of the $SU(2)_L$ interactions can overcome the kinematic effects. Likewise, the chargino preferentially decays into $\tilde{\tau}_2$ rather than $\tilde{\tau}_1$ once $\tilde{\tau}_2$ becomes light enough for a bigger $M_N$ even if $\tilde{\tau}_2$ is still heavier than $\tilde{\tau}_1$. As $M_N$ increases even further, the $\tilde{\tau}_1$ obtains more left-handed component because of the mass suppression of the left-handed stau due to a large neutrino Yukawa coupling. Therefore the decay branching ratio into $\tilde{\tau}_1$ increases while that into $\tau_2$ decreases. The decay into sneutrino still dominates over that into stau for this stau coannihilation region point partly because the right-handed stau is still lighter than the 
left-handed stau doublet as illustrated in Fig.~\ref{rgefig}. Note that, even though $\tilde{l}_R$ is lighter than $\tilde{l}_l$, the branching ratio into $\tilde{l}_R$ is negligible compared with $\tilde{l}_l$ because ${\chi}_1^{+}$-$\tilde{l}_R$ coupling is mainly via the small Higgsino component in the chargino. We also note that, because both $\chi^0_2$ and $\chi^+_1$ are
wino-like gauginos, the analogous arguments apply for the $\chi^0_2$ decays to result in the tau event enhancement.

\subsection{Tau Lepton Signatures at LHC}
\label{eve}
    \begin{figure}[t]
        \centering
                   \includegraphics[width=0.47\textwidth]{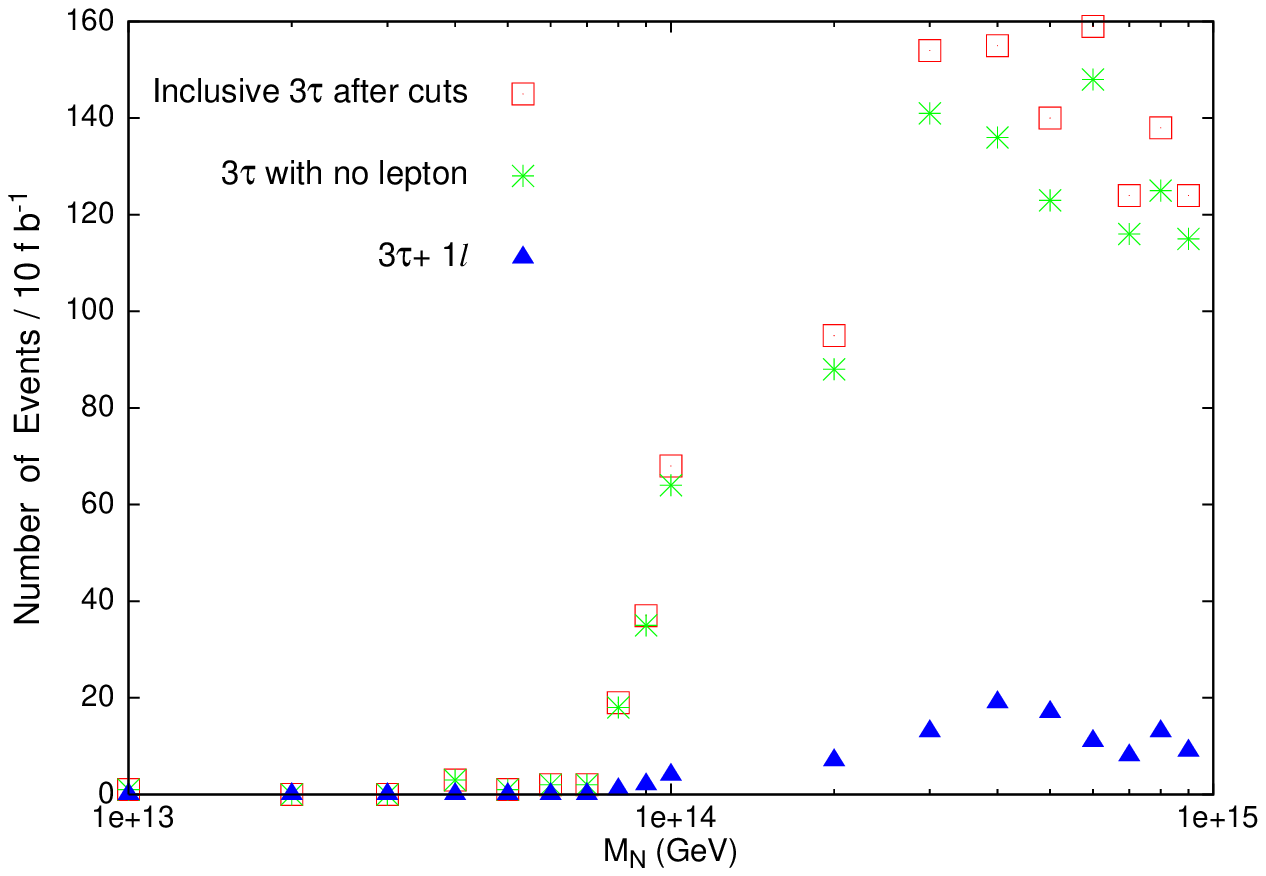}
                 \includegraphics[width=0.47\textwidth]{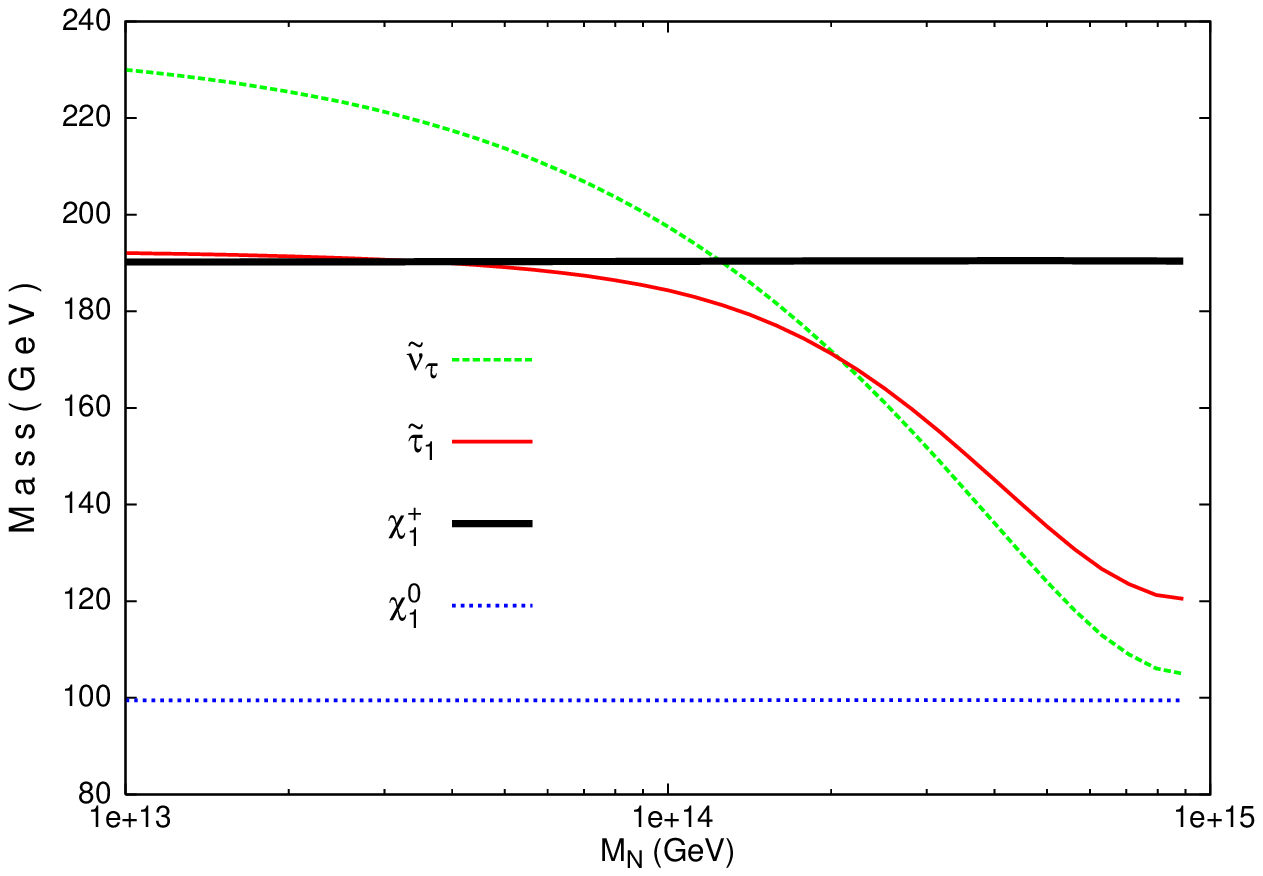}
  \caption{The number of 3 hadronic tau signal events after the cuts as 
a function of the right-handed neutrino mass $M_N~$(GeV) at the 
integrated luminosity of $10 fb^{-1}$. The other parameters are same as those of Point A in Table \ref{bench1}. The sneutrino coannihilation region corresponds to $M_N\sim 8 \times 10^{14}~$GeV and the SM backgrounds are negligible with our cuts. Also shown on the right are the masses of the tau sneutrino, lighter stau, lightest chargino and lightest neutralino.}
  \label{number} 
   \end{figure}
Given the above analysis, one can expect the enhanced tau events when the stau doublet is suppressed.
Let us now verify that such enhancements can be indeed detectable at a collider, most interestingly at the forthcoming LHC, 
for our benchmark points by performing the simple event generations.     
The events were generated by PYTHIA with the center of mass energy $\sqrt{s}=14~$TeV.
We are interested in the sneutrino and stau coannihilation regions in the 
relatively small $m_0,m_{1/2}$ parameter space, and the light gluino and squarks ($\lesssim 1~$TeV) 
are the dominant SUSY particle productions at LHC whose cascade decays lead to multiple jets 
and missing energy. The events are selected by the following set of cuts
\begin{itemize}
\item 
Four jets with $p_T>50~$GeV with the leading jet with $p_T>100~$GeV. 
\item  
The missing transverse energy $\not\!\! E_T$ is required to be
\ba
\not\!\! E_T > max(0.2 M_{eff},100~\mbox{GeV})
\ea
The effective mass $M_{eff}$ is defined as the sum of the missing transverse energy and the 
transverse momenta of the four hardest jets. 
\item
Leptons ($e,\mu$) are required to have $p_T>20~$GeV and $|\eta|<2.5$.
\item
 Hadronically decaying taus are required to have $p_T>20~$GeV and $|\eta|<2.5$. We choose the tau efficiency factor $\epsilon_{\tau}=0.4$ for definiteness.
\end{itemize}
 The leptons are 
isolated by demanding no more than 10 GeV of the transverse energy should be present in a 
cone with a size $R=0.2$ around the lepton. We assume for simplicity the isolated leptons are identified correctly. Jets are identified by a simple cone algorithm PYCELL subroutine of PYTHIA. The calorimeter cells with a uniform segmentation $\Delta \phi=0.1,\Delta \eta=0.1$ for $|\eta|<3$ were used and the cells 
with more than 1 GeV were taken as a possible jet trigger. $R=0.4$ was used as a jet cone size and a cluster was accepted as a jet if the transverse energy sum is bigger than 20 GeV. 

The identification of the hadronic taus heavily depends on the details of the detector performance \cite{aad,cms} and going through the detailed tau identification algorithms is beyond the scope of the paper. To demonstrate the potential significance of the multiple $\tau$ events, we estimate the number of the single-prong and three-prong tau events from the generator information instead of a crude 
estimation by finding a collimated jet around a parent tau without a full detector simulation (i.e. we counted the number of taus before showering which eventually decayed (after showering) to 1 or 3 charged pions). Even though $\epsilon_{\tau}$ can well be bigger than $0.4$ at LHC, we choose a conservative value for a high jet rejection rate \cite{aad,cms}.  For the jet rejection factor (which depends on the jet $E_{T}$ and the tau tagging efficiency $\epsilon_{\tau}$), we use the following values: $300$ for $20<E_{T}<30$, $500$ for $30<E_{T}<60$, $1000$ for  
$60<E_{T}<100$ and $3000$ for $100<E_{T}$ \cite{aad}. QCD jet events have large cross sections so that they are expected to be clarified by the actual testing of several identification/reconstruction algorithms on the first set (integrated luminosity of $\sim 100 pb^{-1}$) of the LHC data. Not to overestimate our tau signal estimations, 
we recorded the fake taus from QCD jets only for the SM backgrounds and not for the SUSY signal events.

For the SM backgrounds, we generated the events for $ZZ,WZ,Z+$Jets$,WW,t\bar{t}$ and checked that they 
are sufficiently reduced by our aforementioned cuts. We generated $28\times 10^7$ events for Z+$Jets$ and $10^7$ events for the other processes \footnote{These numbers of events correspond to the integrated luminosity of about $10 fb^{-1}$ for Z+$Jets$, $20 fb^{-1}$ for  $t\bar{t}$, $140 fb^{-1}$ for WW, $400 fb^{-1}$ for WZ and $1 pb^{-1}$ for ZZ.}. Because of our strong $\not\!\! E_T$ cut along with our tau tagging efficiency $\epsilon_{\tau}=0.4$ and consequently a decent jet rejection rate, we found these standard model backgrounds are negligible for 3 and 4 tau events at the integrated luminosity of our interest $\sim 10 fb^{-1}$. Even though the complete 
background analysis including optimization of the cuts is beyond the scope of this paper, we indeed can verify the main goal of this paper in the following, namely, the big enhancements of the tau signals at LHC in the seesaw scenario even with these strong cuts. \footnote{The backgrounds from the heavy flavors such as 
$Z b \bar{b}$ are also expected to be small because of the suppressions of tau events from a heavy flavor. Even though a more complete analysis for the SM backgrounds including QCD multi-jets is beyond the scope of this paper, we expect that our strong cuts would be sufficient not to affect our main results for the enhancement of the observable multiple ($\geq 3$) tau signals. For instance, multiple fake tau events from QCD processes would be adequately suppressed with our choices of the cuts and rejection rates \cite{aad}. We also note that some preliminary work shows the possibility for the 
much better QCD jet rejection rates than our chosen values \cite{lai}.}

Let us now see the estimations of the tau signal events. The significant enhancements in the observable 3 and 4 tau events were indeed found. For Point A representing the sneutrino coannihilation regions, we found 138 3-$\tau$ events for the seesaw model and only 1 3-$\tau$ event for CMSSM at the integrated luminosity of $10 fb^{-1}$. Hence it can be inferred that such 3-$\tau$ events in the seesaw scenario can be discovered even at the integrated luminosity of order $1 fb^{-1}$. We also found 5 4-$\tau$ events at $25 fb^{-1}$ for the seesaw model at this benchmark point. This indicates that even 4-$\tau$ events, possibly by optimizing the cuts, could be within the kinematical reach at LHC to discover the deviations from the SM predictions. The left-panel of Fig.~\ref{number} shows the enhancement of 3-$\tau$ events 
as a function of the right-handed neutrino mass $M_N$. The non-monotonic behavior of the number of events for a large $M_N$ region (say $M_N \gtrsim 3\times 10^{14}~$GeV) is partly because of the relative mass differences among the light chargino/neutralino and the stau doublet (shown in the right panel) which affect the number of taus passing the cuts. 
We can see $3 \tau + 1 l ~(l=e \mbox{ or }\mu)$ signals could also be within the discovery reach at LHC while the dominant fraction of the 3 tau events include no lepton which are also clean signals.   

Even though Point B has the smaller inclusive cross section because of a larger $m_{1/2}$ and a large branching ratios of charginos and neutralinos into taus even in CMSSM, we indeed found the expected enhancement giving 9 and 24 3-$\tau$ events in the seesaw model respectively at the integrated luminosity of $10 fb^{-1}$ and $20fb^{-1}$ compared with 5 and 8 3-$\tau$ events for the CMSSM.

Such enhancement of the tau events in a minimal extension of the CMSSM would greatly enlarge the parameter space which can be probed at LHC through $\tau$ signals. 

\begin{table}
\begin{minipage}
{0.5\linewidth}\centering
\begin{tabular}{|c|c|c|}
 \hline  
 Point A & Seesaw & CMSSM \\ \hline 
$3 {\tau} $ &  138 &  1 \\ \hline
$2 {\tau} +l$ &  521 &  5 \\ \hline
 $1 {\tau}+2l$  & 468 &  14 \\ \hline
 $ 3 l$  & 109 &  73 \\ \hline
\end{tabular}
\end{minipage}
\hspace{0.5cm}
\begin{minipage}
{0.5\linewidth}
\centering
\begin{tabular}{|c|c|c|}
 \hline  
 Point B & Seesaw & CMSSM \\ \hline 
$3 {\tau} $ &  9 &  5 \\ \hline
$2 {\tau} +l $ &  39  &  18 \\ \hline
 $1{\tau}+2l$  & 57 &  52 \\ \hline
 $3 l$  & 30 &  89 \\ \hline
\end{tabular}
\end{minipage}
\caption{The number of inclusive hadronic tau and lepton signal events at the integrated luminosity of $10 fb^{-1}$. The fake tau signal contributions are not included.}
\label{12tau}
\end{table}

\section{Discussion}
\label{dis}
Before concluding our discussions on the 3 and 4 tau signals at LHC in the seesaw scenario, let us give some comments on other tau channels. Even though we focused on the 3 or more hadronic tau events because they have not received due attention in the literature and hence they are the main results of our paper, the events with a smaller number of taus are also expected to be enhanced in a light stau doublet scenario. Those signals with 2 or less taus with leptons which can also give clean signatures at LHC have been studied extensively and we refer the readers to the existing literature for more details such as the discussions on the SM backgrounds \cite{nojidrees,bae2,drees,kao2,joe,james,matchepie,mat1,hinch2,mre,hinch,dark1,nonzeroA,nonzeroA2,bha}. We here just remark on those tau signal enhancements in the seesaw scenario at 
our benchmark points which are shown in Table \ref{12tau}. The increase in the 
hadronic tau signals in the seesaw scenario compared with CMSSM case is again verified here. For CMSSM, we find less signal events when the signals include more taus. This is reasonable because the effects of the small tau efficiency would suppress the signals 
significantly unless the tau production processes are much larger than the lepton production processes. 
For the seesaw scenario, because we are considering the benchmark points where the tau productions are significantly bigger than the lepton productions, $1\tau +2l$ 
signals can be bigger than $3 l$ signals. As the signals include more taus, however, the signal suppressions due to the small tau tagging efficiency become more important to cancel off the effects by the tau production enhancement. This is reflected in the decline of the signals with more than 2 taus in Point A and more than 1 tau in Point B. We also note that tri-lepton events are indeed suppressed for the seesaw scenario compared    
with CMSSM case at Point B, while the increase of tri-lepton signals for Point A in the seesaw scenario is partly due to the leptons
from the tau decays.
\\
\\
We have studied the collider signatures of the light left-handed stau doublet due to a large neutrino Yukawa coupling in the type-I seesaw scheme. Due to their smaller masses and $SU(2)_L$ interactions, we showed that many tau events can be expected at LHC in the regions with relatively small $m_0,m_{1/2}$. In particular, we showed the signals with $3$ or more taus $+$ jets can be observable at LHC with the integrated luminosity $1 \sim 30 fb^{-1}$ with negligible SM backgrounds. This has been demonstrated through the concrete parameter sets in the sneutrino and stau coannihilation regions with a relatively small $\tan \beta$. This is in contrast to the conventional light stau scenarios in the previous literature which use a large $\tan \beta$ to suppresses the stau mass by a large mixing in the stau mass matrix. The seesaw scenario can realize the light (left-handed) stau by the effects of RGE of the slepton doublet as a consequence of a large neutrino Yukawa coupling. 

We also noted that the kinematically favorable small $m_0,m_{1/2}$ regions with the desirable thermal relic abundance are typically disfavored for a large $\tan \beta$ due to, for instance, the constraints from $Br(b \rightarrow s \gamma)$ (which is the case for both CMSSM and constrained seesaw scenario). Hence the realization of the light stau with a small $\tan \beta$ as we have discussed in the seesaw scenario would be of great interest for the kinematical reach of the search for new physics at LHC in the parameter space compatible with the thermal relic abundance indicated by WMAP. The enhancement of the tau signatures in a minimal extension of the CMSSM would greatly enlarge the parameter space which can be probed at LHC through tau signals. 

While our main purpose was to give an exploratory survey of the potential significance of the tau events for a light left-handed stau scenario, more extensive event analysis including the full detector effects and the tau identification/reconstruction algorithms along with the thorough SM background estimation would deserve further studies. For instance, the properties of the gauginos and stau can be obtained by studying the di-tau invariant mass distribution and the energy distribution among tau decay products \cite{hinch2,hinch,dark1,dark2,noji2,nat,choi,gra,guc}. Even though the lepton search has been more often discussed than the tau search 
partly because taus are harder to identify than electron/muons, tau physics would certainly extend the discovery potentials at LHC in conjunction with the conventional lepton event studies, and more importantly provide a unique way of probing the structure of the possible new physics.
\\
\\
We thank G. Kane, A. Pierce and J. Wells for the useful discussions. This work in part was supported by Michigan Center for Theoretical Physics and the Syracuse University College of Arts and Sciences.


\end{document}